\pdfoutput=1
\documentclass[aps,pre,twocolumn]{revtex4}
\usepackage{graphicx}
\usepackage{amssymb,amsfonts,amsmath}
\usepackage{hyperref}
\usepackage{longtable}
\setlongtables

\begin{document}

\title{Self-Assembly of Patchy Particles into Diamond Structures through Molecular Mimicry}

\author{Zhenli Zhang$^{1}$}
\author{Aaron S. Keys$^{1}$}
\author{Ting Chen$^{1}$}
\author{Sharon C. Glotzer$^{1, 2}$}
\email[Corresponding author.  E-mail: ]{sglotzer@umich.edu}
\affiliation{$^1$ Department of Chemical Engineering and $^2$ Department of Materials Science Engineering, University of Michigan, Ann Arbor MI, 48109-2136}

%\date{\today}

\begin{abstract}
Fabrication of diamond structures by self-assembly is a fundamental challenge in making three-dimensional photonic crystals. We simulate a system of model hard particles with attractive patches and show that they can self-assemble into a diamond structure from an initially disordered state. We quantify the extent to which the formation of the diamond structure can be facilitated by ``seeding'' the system with small diamond crystallites or by introducing a rotation interaction to mimic a carbon-carbon antibonding interaction. Our results suggest patchy particles may serve as colloidal ``atoms'' and ``molecules'' for the bottom-up self-assembly of three-dimensional crystals. 
\end{abstract}

\maketitle

\section{Introduction}

The diamond structure is one of the most desirable structures from which to make photonic crystals because it provides a three-dimensional, complete photonic band gap that allows the crystal to diffract light efficiently\cite{r1}. The current methods of fabrication of these structures involve direct drilling\cite{r2} and layer-by-layer lithography\cite{r3}, which are both top-down approaches and are usually expensive and inefficient. Thus the current fundamental challenge to the materials community is how to fabricate the diamond photonic-band gap structure in an economically feasible and controllable way that is capable of being scaled up to industrial scales. Bottom-up self-assembly is a promising strategy, given the right interactions among the building blocks. There has been substantial interest in assembling 3D crystals from colloidal particles\cite{r4, r5, r6, r7, r8, r9, r10}. However, self-assembling a diamond crystal using a one-component system of colloids has not yet been achieved\cite{r11}. Decorating the surface of colloids with attractive ``patches'' suggests a promising approach to assemble more complex structures, in a potentially controllable and predictable way due to the precise interactions between the patches\cite{r12}.  Many natural and synthetic molecules and particles, such as protein capsomers in virus shells and nanoparticles with binding ligands\cite{r13}, can be viewed as patchy particles which may serve as programmable building blocks for tomorrow's materials\cite{r14}.

Here we propose a model system of patchy particles that, soon, may be possible to fabricate based on recent experimental findings and theoretical analysis~\cite{r15,r16}. By performing Monte Carlo simulations, we show that these model patchy particles are capable of assembling from a disordered state into a diamond lattice, and we quantify the extent to which seeding the system, or adding a rotational interaction, can dramatically facilitate the formation of diamond structured assemblies. 

\subsection{Model and Simulation Method} 

\begin{figure}
\label{fig:fig1}
\centerline{\includegraphics[width=0.75\columnwidth]{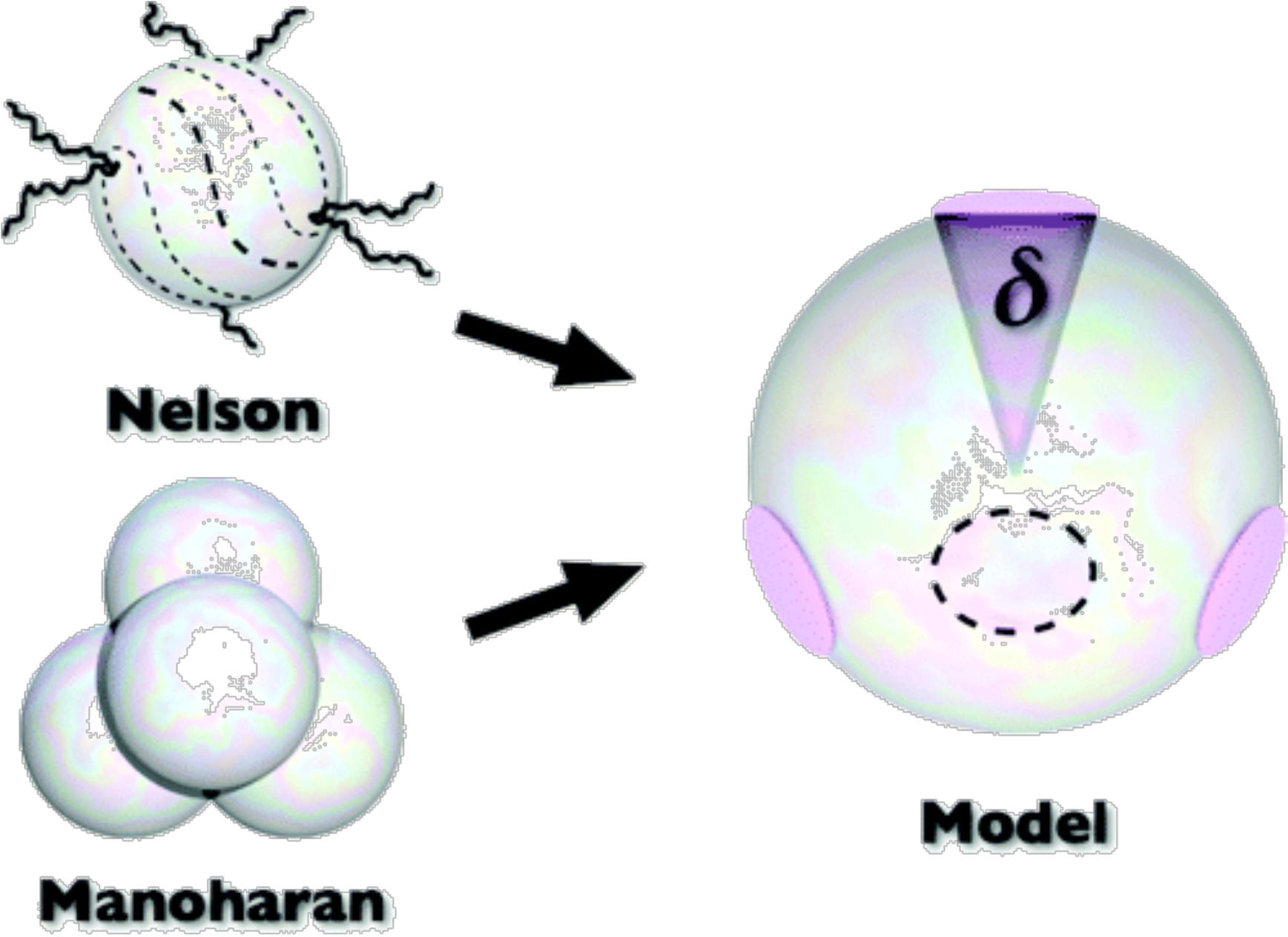}}
\caption{\label{fig:fig1} Model of patchy particle in our simulations (right). Sticky ``patches'' are shown in pink. Dotted circle indicates fourth sticky patch on far side of particle. Related models described in text (left). Nelson model in upper left based on similar schematic in ref~\cite{r15}.}
\end{figure}

The model system we consider is illustrated in Figure 1 (right), which shows a spherical particle with four circular patches arranged in a tetrahedron on the particle surface. A pairwise attraction is assigned between patches on different particles to allow the particles to associate with each other so as to reproduce tetrahedral symmetry in three spatial dimensions. The possibility of synthesizing this building block experimentally is supported by two recent reports. The first is the theoretical analysis by Nelson\cite{r15}, who proposed a method to make tetravalent colloids in which micron-scale spherical colloids are coated by anistropic nanometer-sized objects. The ordering of the objects on the particle surface produces four disclination defects that may allow the creation of tetravalent colloids with chemical or biomolecule linkers such as DNA oligonucleotides anchored at the defect cores, as shown in Figure 1 (upper left). An alternative approach is provided by tetrahedral clusters as in Figure 1(lower left) formed from micron-sized spheres via evaporation-driven self-assembly of colloids\cite{r16}.  By backfilling these ``secondary building blocks'', it has been proposed to make spheres in which each of the four colloids within the cluster protrudes slightly beyond the radius of the outer sphere, providing circular ``patches'' of material different from that of the outer sphere. To first approximation, both types of building blocks can be modeled by a sphere with four circular patches arranged tetrahedrally.

The spherical shape of our model patchy particle allows the use of a pair potential between patches expressed as a hard sphere, square-well potential modulated by an angular term\cite{r17} 
\begin{equation}
U_{ij}(r_{ij}; \mathbf{Q}_i, \mathbf{Q}_j) = u^{\mathrm{hssw}} (r_{ij}) \cdot f( \mathbf{Q}_i, \mathbf{Q}_j),
\end{equation}
where $u^{\mathrm{hssw}}$ is the regular hard sphere square well potential with reduced range
\begin{equation}
u^{\mathrm{hssw}}(r_{ij}) = 
\begin{cases}
\infty & \mathrm{for}\quad  r < \sigma \\
-\epsilon & \mathrm{for}\quad \sigma \leq r < \lambda \sigma \\
0 & \mathrm{for}\quad r \geq \lambda \sigma
\end{cases}
\end{equation}
Here $\sigma$ is the diameter of the particles and set to be unity, and $\epsilon$ is the depth of the square well potential. The function $f(\mathbf{Q}_i, \mathbf{Q}_j) = 1$ only if particles $i$ and $j$ are oriented so that the vector joining their mass centers intersects an attractive patch on both particles; otherwise $f(\mathbf{Q}_i, \mathbf{Q}_j) = 0$. The angle $\delta$, shown in Figure 1(right), represents the patch size, and $\lambda$ represents the interaction range. The two values in combination determine the directionality and preciseness of the interaction between any two patches. In this study we fix the location of the patches and choose $\delta = \pi/6$, $\lambda = 1.1\sigma$ which, for a $1 \mu m$ diameter particle corresponds to a patch size of $0.26 \mu m$ and an interaction range of $0.1\mu m$. We leave the effect of size, interaction range and disorder in the location of the patches for future investigation.

We perform Monte Carlo simulations using this model in a cubic box of fixed size with periodic boundary conditions. Since we are interested in the diamond structure (density $\rho_\mathrm{d} = 0.6495$ for a perfect, box-spanning, diamond crystal), we limit our study to $\rho < \rho_\mathrm{d}$. All simulation runs begin from a disordered state at high temperature $T$ and are subsequently cooled to the target temperature. We investigate three types of systems: systems without small diamond crystal seeds, systems with seeds, and systems with a modified potential to induce the rotation of particles relative to ``bonded'' neighbors. The details of the potential used for the third system will be discussed later.

\subsection{Results and Discussion}

\subsubsection{Systems Without Seeds}

For the unseeded system, we examine the effect of two different cooling rates on assembly. In the ``fast'' run, $T$ is decreased by $\Delta T = 0.001$ over cycles of 1 million Monte Carlo steps (MCS), where $T$ is defined in units of $\epsilon/k_\mathrm{B}$.  A MC step is defined here as one attempted MC move per particle. In the ``slow'' run,$ T$ is decreased by $\Delta T = 0.001$ over cycles of 10 million MCS. A Monte Carlo move for a randomly selected particle is randomly chosen to be either a translation or a rotation with equal probability. The maximum values of the translation and rotation are variable to keep the acceptance ratio at 0.5. Depending on the density, the maximum translation varies from 0.02 to 0.04 particle diameters and the maximum rotation ranges from 0.05 to 0.09 rad. Although there is no simple and direct connection between the number of Monte Carlo steps and the physically relevant time scales of the system, the purely local, physical moves of the simulation provide a reasonable approximation to local diffusive motion. Eleven or twelve independent runs are performed to obtain the statistical results shown for each density. For the fast cooling rate, only kinetically arrested, disordered structures are obtained. However, at the slow cooling rate we observe the formation of diamond structures in some of the runs. The fraction of successfully assembled diamond structures is 55\% (six times in 11 runs) at $\rho = 0.40$. A typical structure obtained at $\rho = 0.40$ is shown in Figure 2a. To characterize the structure we calculate the pair-correlation function, $g(r)$, shown in Figure 2b. The correlation function exhibits peaks at $\sim1.0\sigma$, $\sim1.633\sigma$, and $\sim1.954\sigma$, which are the locations of the first three characteristic peaks of a perfect diamond crystal. The bond angle distribution function, $g(\theta)$, is shown in Figure 2c. We observe two pronounced peaks at $\theta \approx 60^{\circ}$ and $\theta \approx 180^{\circ}$, which together with $g(r)$ confirms the diamond structure for the slow cooling rate.

\begin{figure}
\label{fig:fig2}
\centerline{\includegraphics[width=0.75\columnwidth]{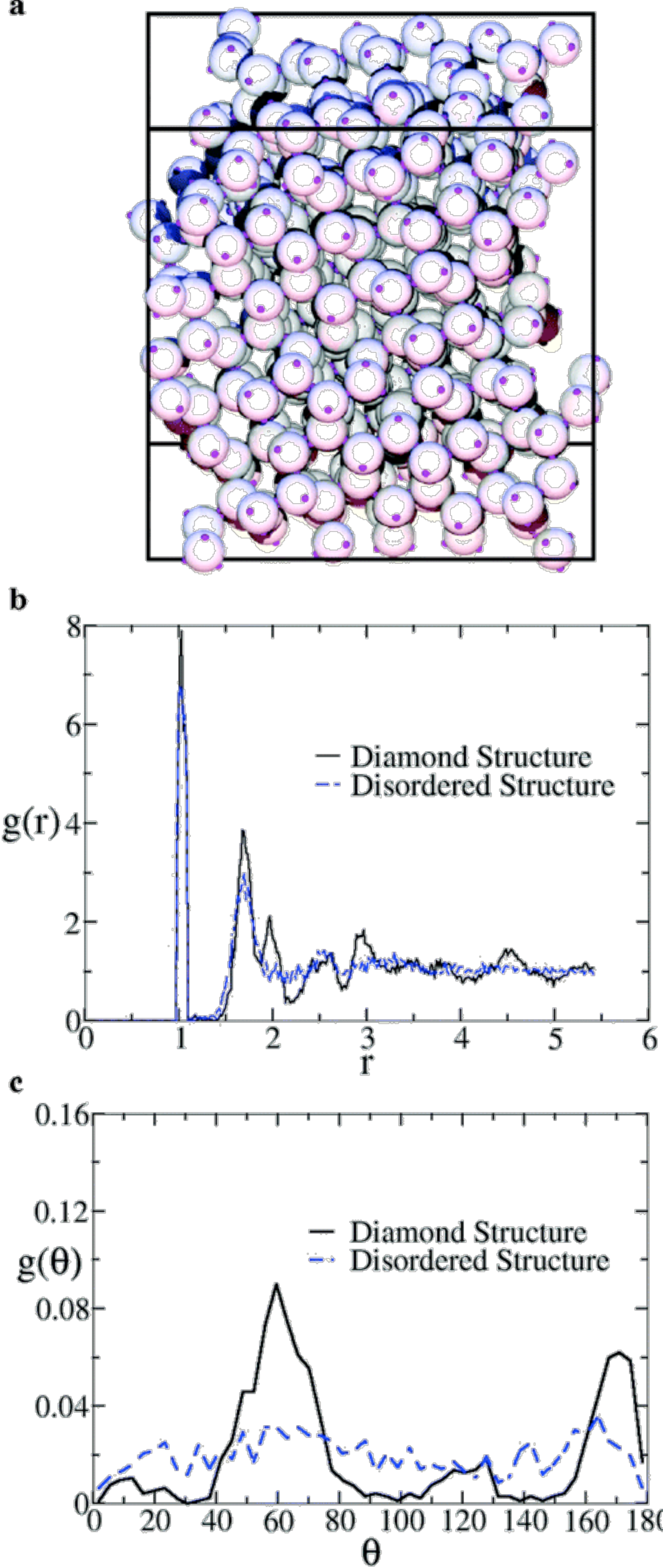}}
\caption{\label{fig:fig2} Simulation results of the system without seeds. (a) Diamond structure obtained at slow cooling rate for $\rho = 0.40$, $T = 0.11$. (b) Pair correlation function $g(r)$ for disordered structures obtained with fast cooling rate and diamond structures obtained with slow cooling rate. Note absence of characteristic third peak for disordered structure. (c) Bond-angle distribution $g(\theta)$ for the diamond and disordered structures.}
\end{figure}

We calculate the degree of crystallinity $\varphi_c$ and a bond-angle correlation function $C_\mathrm{b}$ to further quantitatively characterize the crystal structures. We define $\varphi_c = N_\mathrm{s}/N$ where $N_\mathrm{s}$ is the number of ``solidlike'' particles, which we define as particles in a locally diamond-like configuration. To identify the solid particles, we calculate a local order parameter $q_{3m}(i)$ based on spherical harmonics $Y_{3m}(r_{ij})$ \cite{r18} 
\begin{equation}
\bar{q}_{3m}(i) = \frac{1}{N_\mathrm{b}} \sum_{j=1}^{N_\mathrm{b}(i)} Y_{3m}(\theta(\mathbf{r}_{ij}, \phi(\mathbf{r}_{ij}),
\end{equation}
\begin{equation}
\mathbf{q}_{3}(i) = \frac{\bar{q}_{3m}(i)}{\left( \sum_{m=-3}^{3} | \bar{q}_{3m}(i) | \right)^{1/2} } ,
\end{equation}
where $\mathbf{q}_3$ is a complex vector.  If the number of neighboring particles with which a given particle $i$ is ``bonded'' exceeds three, and if the dot product of $\mathbf{q}_3(i)$ and $\mathbf{q}_3(j)$ is less than 0.65, particle $i$ is identified as a solid-like particle.  The bond-angle correlation function $C_\mathrm{b}$ is expressed by
\begin{equation}
C_\mathrm{b} = - \left< \mathbf{q}_3(i) \cdot \mathbf{q}_3(j)^* \right>
\end{equation}
 We find $\phi_\mathrm{c} =$ 68.9\% and $C_\mathrm{b} = 0.779$ for the imperfect diamond structure in Figure 2a, as compared to corresponding ideal values of 100\% and 1.0, respectively, for a perfect diamond crystal. The calculated values are much higher than those of the disordered structures, as shown in Figure 3, which includes data from several densities.

\begin{figure}
\label{fig:fig3}
\centerline{\includegraphics[width=0.75\columnwidth]{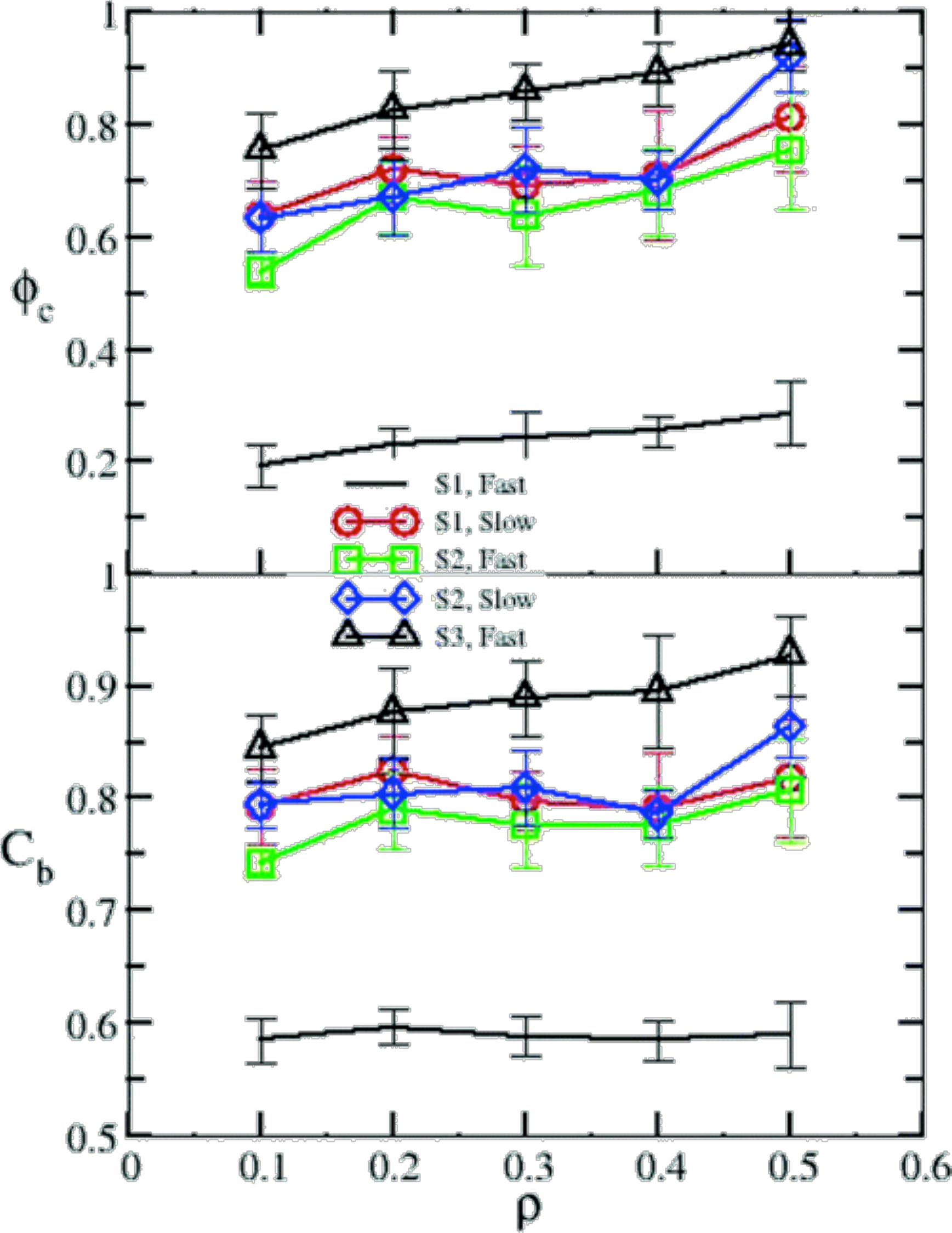}}
\caption{\label{fig:fig3} Structural information for all systems. S1 corresponds to system without seeds, S2 corresponds to the seeded system, and S3 corresponds to the system with ``antibonding'' rotational interactions. (Top) Degree of crystallinity $\varphi_\mathrm{c}$. (Bottom) Bond-angle correlation function $C_\mathrm{b}$.}
\end{figure}

Figure 4a shows the time evolution of $\varphi_\mathrm{c}$ and $C_\mathrm{b}$ for the slowly cooled system in Figure 2a. We see that both quantities consistently show a sharp transition near $T = 0.133$, which indicates the occurrence of a structural transition. In contrast, no obvious change in the corresponding curves is found in the systems formed via the fast cooling rate. A plot of the ``phase boundary'' between diamond and disordered structures determined by this method is presented in Figure 4b.

\begin{figure}
\label{fig:fig4}
\centerline{\includegraphics[width=0.75\columnwidth]{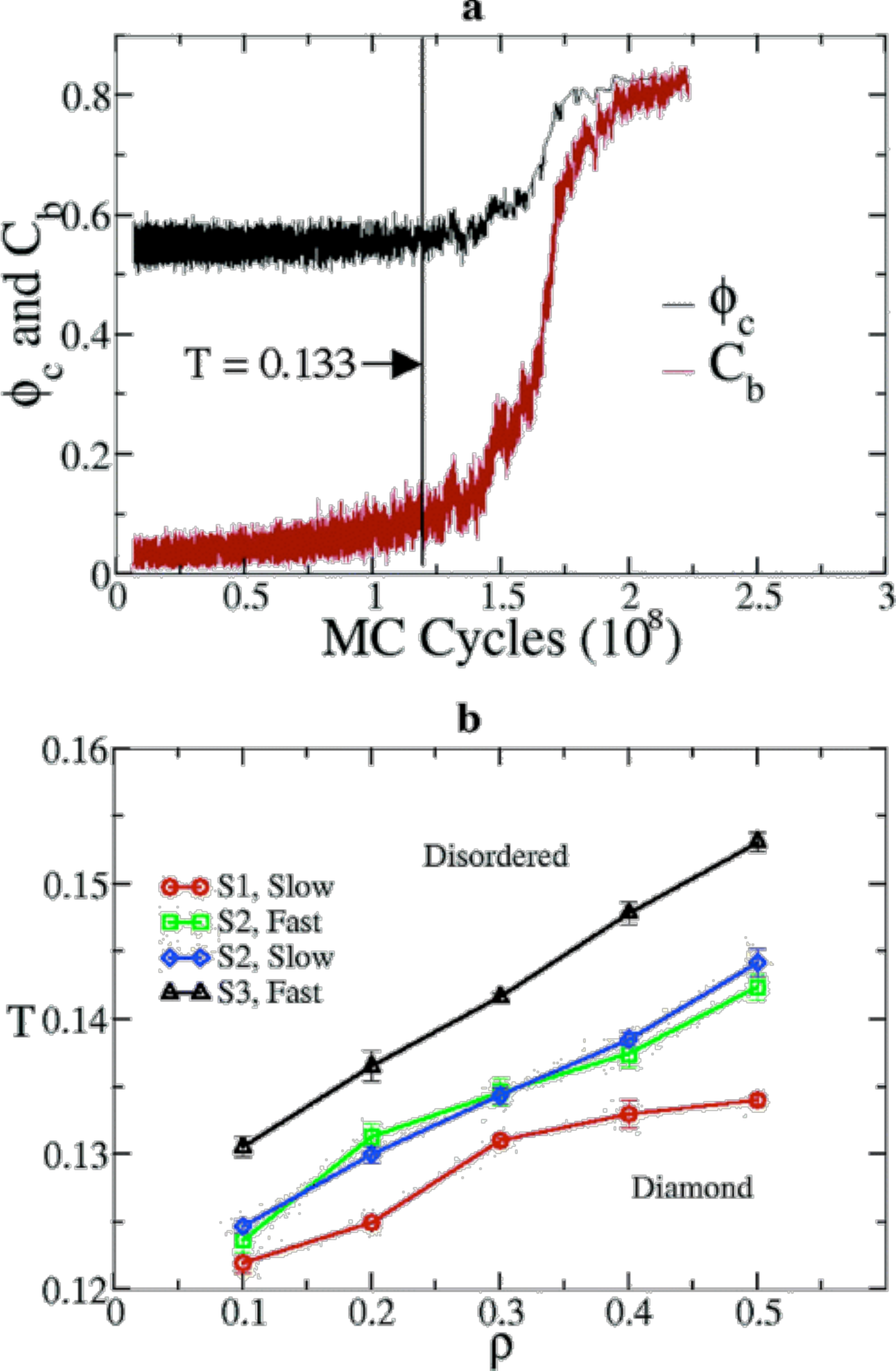}}
\caption{\label{fig:fig4} (a) Evolution of degree of crystallinity $\varphi_\mathrm{c}$ and bond-angle correlation function $C_\mathrm{b}$ for the system in Figure 2a. (b) Plot of maximum $T$ at which diamond structure forms at a given density $\rho$ for systems S1, S2 and S3 as defined in the caption of Figure 3.}
\end{figure}

\subsubsection{Seeded Systems}

It is well-known that introducing small crystalline seeds of the desired structure into a liquid can accelerate crystallization in molecular and colloidal systems\cite{r19}.  Therefore, it is of interest to ascertain to what extent this approach facilitates the formation of the desired diamond structure. To introduce the seeds, we freeze eight of the 512 particles in the simulation box into a single unit cell of a perfect diamond structure and let them remain immobile throughout the simulation. Prior to cooling, all other particles except the seed particles are randomly distributed in the simulation box. We investigate the same two cooling rates as for the unseeded system.

%Table 1.? Probability of Observing Diamond Structure for Various Systemsa
% 	fast$\varphi_\mathrm{c}$ooling?rate	slow$\varphi_\mathrm{c}$ooling?rate
% 	S1	S2	S3	S1	S2
%$\rho=$0.10 	0%(0/12) 	25%(3/12) 	75%(9/12) 	25%(3/12) 	58%(7/12)
%$\rho=$0.20 	0%(0/12) 	42%(5/12) 	100%(11/11) 	25%(3/12) 	50%(6/12)
%$\rho=$0.30 	0%(0/12) 	33%(4/12) 	100%(12/12) 	50%(6/12) 	75%(9/12)
%$\rho=$0.40 	0%(0/12) 	64%(7/11) 	100%(12/12) 	55%(6/11) 	64%(7/11)
%$\rho=$0.50 	0%(0/12) 	50%(6/12) 	100%(12/12) 	75%(9/12) 	67%(8/12)
%
%a?S1, S2, and S3 are defined in the caption of Figure 3. The percentages in the table correspond to Nd/Nr ? 100%, where Nd = number of runs in which diamond structure is observed. Nr = total number of total runs. Nd and Nr are indicated in parentheses.

\begin{table*}
\caption{\label{table:param} Probability of Observing Diamond Structure for Various Systems.$^a$ }
\begin{tabular}{| c | c c c | c c |}
\hline   
\ & 
\multicolumn{3}{c |}{Fast cooling rate} &
\multicolumn{2}{c |}{Slow cooling rate}  \\
%\multicolumn{2}{c}{Slow cooling rate} \\ 
\hline
\ & S1 & S2 & S3 & S1 & S2 \\
 \hline
$\rho=$0.10 &	0\%(0/12) 	& 25\%(3/12) &	75\%(9/12) &	25\%(3/12) &	58\%(7/12) \\
$\rho=$0.20 &	0\%(0/12) 	& 42\%(5/12) &	100\%(11/11) &	25\%(3/12) &	50\%(6/12) \\
$\rho=$0.30 &	0\%(0/12) 	& 33\%(4/12) &	100\%(12/12) &	50\%(6/12) &	75\%(9/12) \\
$\rho=$0.40 &	0\%(0/12) 	& 64\%(7/11) &	100\%(12/12) &	55\%(6/11) &	64\%(7/11) \\
$\rho=$0.50 &	0\%(0/12) 	& 50\%(6/12) &	100\%(12/12) &	75\%(9/12) &	67\%(8/12) \\
\hline
\end{tabular}
\begin{minipage}[t]{0.6\textwidth}
\footnotetext{$^a$S1, S2, and S3 are defined in the caption of Figure 3. The percentages in the table correspond to $N_\mathrm{d}/N_\mathrm{r} \times 100\%$, where $N_\mathrm{d}$ is the number of runs in which diamond structure is observed and $N_\mathrm{r}$ is the total number of total runs. $N_\mathrm{d}$ and $N_\mathrm{d}$ are indicated in parentheses.} 
\end{minipage}
\end{table*}

With the introduction of the seeds, we observe diamond structures for the fast cooling rate, which did not permit crystallization on the time scales of our simulations in the absence of a seed (Table 1). A typical structure at $\rho = 0.40$ is shown in Figure 5a. The fraction of times the diamond structure is obtained in \cite{r11} independent runs at a fast cooling rate increases from 0\% for the systems without seeds to 64\% for the systems with seeds. The same trend is observed for the slow cooling rate; although the fraction increases less dramatically from 55\% to 64\%, the facilitation is still evident. Table 1 shows that seeding facilitates crystallization at nearly all densities and cooling rates. Due to the mismatch between different crystalline domains around the seed, the degree of crystallinity and the bond-angle correlation functions for the seeded systems are, however, not significantly higher than the slowly cooled systems without seeds. 

\begin{figure}
\label{fig:fig5}
\centerline{\includegraphics[width=0.75\columnwidth]{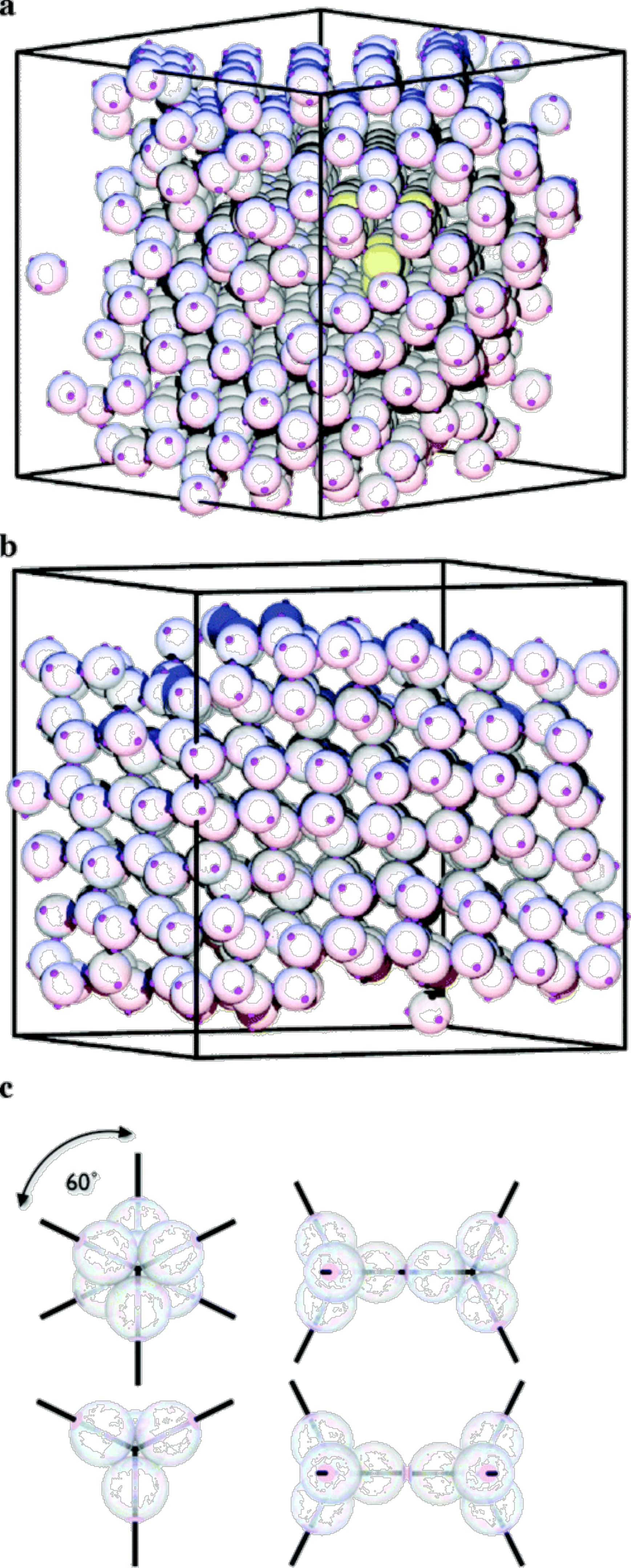}}
\caption{\label{fig:fig5} Structure for the systems with seeds and ``antibonding'' interactions. (a) Diamond structure for the seeded system for $\rho = 0.40$, $T = 0.11$, cooled from T = 0.\cite{r14}5. (b) Diamond structure for the system with rotational interactions, for $\rho = 0.40$, $T = 0.11$. (c) Schematic illustration of two rotation states of ``bonded'' tetrahedrally packed cluster of four spheres. Top: low energy state. Bottom: high energy state.}
\end{figure}

\subsubsection{Systems with Particle-Particle ``Bond'' Rotation}

If we consider the attractive interaction between a pair of patches as a type of ``bonding'', then all of the bonds in our colloidal diamond structure are rotated by $60^\circ$ or $180^\circ$, as are the carbon-carbon bonds in the staggered conformation of ethane. Inspired by this atomistic feature of chemical bonding, we introduce an additional potential energy $u_{ij} = \Delta U \cos(3\theta)$ between pairs of bonded patchy particles, where $\theta$ is the relative rotation angle and $\Delta U \approx 1k_\mathrm{B}T$. This additional potential energy term induces a relative orientation between particles that favors the formation of the diamond structure at low $T$. 

This rotation preference facilitates formation of the diamond structure substantially as compared to the first two methods; this is quantified in Table 1. We find diamond structures for all \cite{r12} independent runs at the fast cooling rate and density $\rho = 0.40$. A typical structure is shown in Figure 5b. Further analysis of $\varphi_\mathrm{c}$ and $C_\mathrm{b}$ (Figure 3) indicates the diamond structures have fewer defects than those obtained by the first two approaches. For example, $\varphi_\mathrm{c}$ = 90.9\% and $C_\mathrm{b} = 0.922$ for $\rho = 0.40$, indicating the formation of a nearly perfect diamond structure.
How might such an interaction be induced between patchy particles?  The rotation of a chemical bond in molecules such as ethane is due to antibonding interactions between the hydrogen atoms on opposite ends of the molecule. Here we offer two possibilities for mimicking this interaction with particles. The first approach involves the use of charges.  Assume the patchy particle modeled here represents a tetrahedral cluster of four spheres fused permanently together.  If the colloidal spheres carry a uniform surface charge, the repulsive electrostatic interaction between spheres in different clusters will induce a similar rotation, as shown schematically in Figure 5c. For $1 \mu m$-diameter colloidal spheres, we estimate that the surface charge needed on each sphere to induce an energy difference between two different rotation states of $1k_\mathrm{B}T$ at $T = 298$K is 0.011231 $C^2/m^2$, based on a model for a solution of charged colloids or proteins in water proposed by Phillies\cite{r20}.  Considering the additional repulsive electrostatic interaction, an attraction energy $\epsilon$ of a patch-patch interaction or ``bond'' of roughly $3.0 \times 10^{-17}$J should be sufficient to induce formation of the diamond structure.

The second approach uses a ``patterned'' patch. In the present work, the patches are considered circular and uniform, but they could be made to be nonuniform like the linear bi-patch pattern investigated computationally in our previous work, which also induces rotation\cite{r12}.  Thus by combining an attractive interaction between patches to provide a ``bond'' between particles with an additional interaction that provides bond rotation, colloidal spheres can be made to mimic tetravalent carbon atoms. It should be noted that the use of seeds and ``antibonding'' interactions apply to thermal systems as well as to systems where cooling rate is not a factor, as in suspensions of large colloids. 

\section{Conclusions}

In summary, our simulations predict that patchy particles with small attractive patches located on the particle surface at the corners of a tetrahedron provide a promising route to achieve a diamond structure through self-assembly.  Introducing a small amount of seeds or an ``antibonding-like'' interaction greatly facilitates the formation of the diamond structure. Our simulations demonstrate that atomic and molecular interactions can guide the design of patchy particles as a new generation of ``colloidal molecules'' \cite{r14, r21, r22} to achieve colloidal crystal analogues of atomic and molecular crystals. 

\begin{acknowledgments}
Financial support was provided by the Department of Energy, Grant No. DE-FG02-02ER46000. We thank M. A. Horsch, J. Mukherjee, M. J. Solomon, and N. A. Kotov for discussions and helpful comments on the manuscript. We thank the University of Michigan Center for Advanced Computing for support of our computing cluster.
\end{acknowledgments}

%\bibliography{preprint}

\end{document}